\documentclass[aps,pra,epsfigure,twocolumn,superscriptaddress,10pt]{revtex4-2}
\usepackage[colorlinks=true,linkcolor=blue,urlcolor=blue,citecolor=blue,pdfusetitle]{hyperref}
\usepackage[utf8]{inputenc}
\usepackage{amsmath}
\usepackage{amsfonts}
\usepackage{bbold}
\usepackage{bbm}
\usepackage{amssymb}
\usepackage{color}
\usepackage{xcolor}
\usepackage{braket}

\usepackage{orcidlink}
\begin{document}

\title{Engineering Atom–Photon Hybridization with Density-Modulated Atomic Ensembles in Coupled Cavities}

\author{Carlos E. M\'aximo}
\email{dumax1@gmail.com}
\affiliation{Institute for Quantum Electronics, ETH Zurich, 8093 Zurich, Switzerland}
\affiliation{Departamento de F\'isica, Universidade Federal de S\~ao Carlos,
	Rodovia Washington Lu\'is, km 235—SP-310, 13565-905 S\~ao Carlos, SP, Brazil}
\author{Romain Bachelard}
\affiliation{Departamento de F\'isica, Universidade Federal de S\~ao Carlos,
	Rodovia Washington Lu\'is, km 235—SP-310, 13565-905 S\~ao Carlos, SP, Brazil}
\author{Tobias Donner}
\affiliation{Department of Physics, ETH Zurich, 8093 Zurich, Switzerland}

\date{\today}

\begin{abstract}
Radiation–matter hybridization allows atoms to serve as mediators of effective interactions between light modes and, conversely, to interact among themselves via light. Here we exploit the spatial structure of atomic ensembles to control the coupling between modes of distinct cavities, thereby reshaping the resulting atom–photon spectra. We show that extended homogeneous clouds suppress mode–mode couplings through destructive interference, whereas grated clouds can preserve them under specific Bragg conditions. This leads to mode-mode spectral subsplittings, where collectivity arises not only from the number of atoms but also from the ability to tune modes of different cavities independently. Our results establish spatially engineered atomic ensembles as a pathway to selective photon transfer between modes and precise control of many-body complexity.
\end{abstract}

\maketitle

\section{Introduction} 

The preparation of atomic ensembles inside high-finesse optical cavities represents a major milestone in cavity quantum electrodynamics (CQED), as it grants access to regimes of collective strong light–matter coupling~\cite{Haroche1983,Agarwal1984,Carmichael1989,Mossberg1990,Zhu2007,Esslinger2007,Zhu2010,Goldwin2016}. Such an achievement stimulated a wide range of ensemble-based applications, from controlling chemical reactions~\cite{Owrutsky2020,Xiong2024} to implementing quantum memories~\cite{Auffeves2011,Faraon2023} and quantum transduction protocols~\cite{Simon2020}. Ideal global coupling, as captured by the Tavis–Cummings model~\cite{Tavis1968}, requires that all atoms couple identically with strength $g$ to a single cavity mode — a condition attainable only when the atoms are indistinguishable, as in a Bose–Einstein condensate~\cite{Esslinger2007}. In practice, even a spatially homogeneous ensemble of $N$ atoms departs from this limit: the spatial averaging over atomic positions and cavity-mode functions effectively reduces the collective coupling from $g\sqrt{N}$ to $g\sqrt{N/2}$~\cite{Birgitta2004,Esslinger2007,Domokos2025}, weakening atom–photon hybridization in the resulting Tavis–Cummings polaritons. More generally, any spatial inhomogeneity in the atomic distribution can degrade collectivity~\cite{Esslinger2013}, with consequences for nonclassical properties such as cavity-field photon statistics~\cite{Solano2007} and bipartite atomic entanglement ~\cite{Retamal2007feb,Retamal2007oct}. Therefore, the nonuniformity of the coupling in CQED systems is not merely a detail, but a key ingredient that modifies fundamental atom–light properties~\cite{Faraon2023}.

In multimode cavity scenarios, atomic ensembles act as mediators of effective intermode interactions, leading to additional avoided crossings in the energy spectra~\cite{Emary2013,Renzoni2013,Zhang2024}. Understanding how density profiles shape such spectral structures is not only of fundamental interest, but also paves the way for new self-organized phases in cold atoms~\cite{Goldbart2009,Donner2017}, controlled photon transfer in cavity arrays~\cite{Aoki22}, and the implementation of quantum gates~\cite{Guo2009,Villasboas2024}. Along these lines, Ref.~\cite{Renzoni2013} reported four-mode coupling in a single Fabry–Perot cavity through atomic clouds with a Gaussian spatial distribution, where both the number of resonances and the magnitude of the splittings were controlled by the cloud width. More recently, Ref.~\cite{Zhang2024} theoretically  investigated two counterpropagating modes in a ring cavity coupled through one-dimensional atomic arrays, showing that the lattice spacing sets the spectral modifications. In these two previous works, however, the atom–mode couplings of the different interacting modes were not independently tunable, a key simplification relative to the framework developed here.

In this work, we show that spatial modulations in the atomic density are crucial to preserve the coupling between modes of distinct cavities, and thereby allow one to manipulate the energy spectrum of many-body CQED systems. Motivated by recent experiments with ultracold atoms in mutually rotated linear cavities~\cite{Donner2017,Donner2018,Donner2025}, we derive a $M$-mode model in which mode-mode couplings are expressed through a density-dependent structure-factor matrix. We demonstrate that large Gaussian clouds suppress this structure factor through destructive interference, while a grating imposed on the atomic profile restores intermode couplings via Bragg scattering. Under such a constructive interference condition, we uncover subsplittings in which collectivity depends not only on the number of atoms but also on the distribution of different atom–mode couplings. Our approach offers a practical means to control the complexity of many-body CQED systems, with potential relevance for multimode quantum technologies~\cite{Moussa2014,Moussa2017,Kryuchkyan2012,Villasboas2024,Rossatto2025}.

The manuscript is organized as follows. Section II introduces the derivation of a coupled-mode model from the master equation, showing how the atomic density profile enters the dynamics through a structure-factor matrix that governs effective mode–mode couplings. Section III analyzes Gaussian atomic clouds, discussing both the subwavelength limit, where uniform collective coupling is recovered, and the large-cloud regime, where destructive interference suppresses intermode scattering. Section IV is devoted to grated atomic distributions, demonstrating how spatial modulation restores intermode coupling in extended clouds via Bragg scattering and leads to additional collective resonances. The Appendixes collect the analytical derivations supporting these results.

\section{Coupled-mode model} 

\begin{figure}
\begin{centering}
\includegraphics[width=0.9\columnwidth]{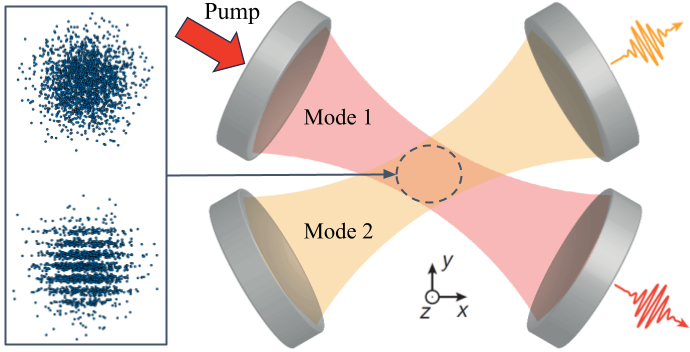}
\par\end{centering}
\caption{Two crossed cavity modes couple to Gaussian atomic distributions of two-level atoms, with and without a grating, in the intersecting region where the light fields combine. A pump field inject photons into mode 1 (in light coral) on the left, while detectors collect the leaking light from the mirrors on the right side.}
\label{fig1}
\end{figure}

We consider an ensemble of $N$ ultracold two-level atoms simultaneously interacting with $M$ quantized light modes, each confined to a different optical cavity. The timescales of interest are such that the internal atomic dynamics is not affected by atomic motion The detuning $\Delta_m$ of each mode $m$ from the atomic transition, as well as the corresponding constant atom-mode coupling $g_m$, can be individually controlled. This system is described by the following Hamiltonian, derived in the rotating-wave approximation:
\begin{multline}
\hat{H}=\eta\left(\hat{a}_{1}^{\dagger}+\hat{a}_{1}\right)+\sum_{m=1}^{M}\left(\Delta_{m}-\Delta\right)\hat{a}_{m}^{\dagger}\hat{a}_{m}-\frac{\Delta}{2}\sum_{j=1}^{N}\hat{\sigma}_{j}^{z}\\+\sum_{j=1}^{N}\sum_{m=1}^{M}g_{m}\cos\left(\mathbf{k}_{m}\cdot\mathbf{r}_{j}\right)\left(\hat{a}_{m}\hat{\sigma}_{j}^{\dagger}+\hat{a}_{m}^{\dagger}\hat{\sigma}_{j}\right), \label{H}
\end{multline}
where the raising (lowering) operator $\hat{\sigma}_{j}=\ket{e_j}\bra{g_j}$ ($\hat{\sigma}_{j}^{\dagger}=\ket{g_j}\bra{e_j}$) of atom $j$ describes the transition from its internal ground state $\ket{g_{j}}$ to its excited state $\ket{e_{j}}$, whereas  $\hat{\sigma}_{j}^z=\ket{e_j}\bra{e_j}-\ket{g_j}\bra{g_j}$ corresponds to the inversion population operator.  The symbol  $\hat{a}_{m}$($\hat{a}_{m}^{\dagger}$) denotes the photon annihilation (creation) operator for mode $m$, which is characterized by its wavevector $\mathbf{k}_{m}$. The standing-wave profile of mode $m$ modulates the atom-mode coupling as $g_{m}\cos(\mathbf{k}_{m}\cdot\mathbf{r}_{j})$, with $\mathbf{r}_{j}$ representing the position of atom $j$. We pump only mode $m=1$ using a laser with Rabi frequency $\eta$ and detuning $\Delta$ from the atomic transition, and adopt it as the reference frame. The remaining $M-1$ modes thus acquire photons solely through scattering by the atomic cloud. Fig.~\ref{fig1} illustrates the particular case of two modes in crossed cavities.

From Hamiltonian~\eqref{H}, we obtain the quantum-optics master equation,
\begin{flalign}
\frac{d\hat{\rho}}{dt}=&-i\bigl[\hat{H},\hat{\rho}\bigr]+\Gamma\sum_{j=1}^{N}\left(\hat{\sigma}_{j}\hat{\rho}\hat{\sigma}_{j}^{\dagger}-\frac{1}{2}\left\{ \hat{\sigma}_{j}^{\dagger}\hat{\sigma}_{j},\hat{\rho}\right\} \right)\nonumber\\&+\sum_{m=1}^{M}\kappa_{m}\left(\hat{a}_{m}\hat{\rho}\hat{a}_{m}^{\dagger}-\frac{1}{2}\left\{ \hat{a}_{m}^{\dagger}\hat{a}_{m},\hat{\rho}\right\} \right), \label{me}
\end{flalign}
where the decay rate $\Gamma$ of the excited atomic state and the dissipation rate $\kappa_{m}$ of each mode $m$ induce a steady state in the system. Eq.~\eqref{me} yields the following coupled equations for the expectation values of atomic and cavity-mode operators:
\begin{eqnarray}
\frac{d\bigl\langle\hat{a}_{m}\bigr\rangle}{dt}&=&\left[i\left(\Delta-\Delta_{m}\right)-\frac{\kappa_{m}}{2}\right]\bigl\langle\hat{a}_{m}\bigr\rangle\nonumber\\&&-i\eta\delta_{m,1}-ig_{m}\sum_{j=1}^{N}\cos\left(\mathbf{k}_{m}\cdot\mathbf{r}_{j}\right)\bigl\langle\hat{\sigma}_{j}\bigr\rangle,\\\frac{d\bigl\langle\hat{\sigma}_{j}\bigr\rangle}{dt}&=&\left(i\Delta-\frac{\Gamma}{2}\right)\bigl\langle\hat{\sigma}_{j}\bigr\rangle\nonumber\\&&+i\sum_{m=1}^{M}g_{m}\cos\left(\mathbf{k}_{m}\cdot\mathbf{r}_{j}\right)\bigl\langle\hat{a}_{m}\hat{\sigma}_{j}^{z}\bigr\rangle. \label{eq:sigma}
\end{eqnarray}
We then focus on the weak-drive limit, $\eta \ll \Gamma, \kappa_m$, where the atomic excited-state population is negligible, $\langle \hat{\sigma}_{j}^{z} \rangle \approx -1$, so that only the coherences remain relevant to describe the atom dynamics. In this regime, the quantum correlations simplify to $\langle \hat{\sigma}_{j}^{z}\hat{a}_{m} \rangle \approx -\langle \hat{a}_{m} \rangle$, closing a set of linear algebraic equations for the steady-state variables:
\begin{eqnarray}
\bigl\langle\hat{a}_{m}\bigr\rangle_{ss}&=&\frac{\eta\delta_{m,1}+g_{m}\sum_{j=1}^{N}\cos\left(\mathbf{k}_{m}\cdot\mathbf{r}_{j}\right)\bigl\langle\hat{\sigma}_{j}\bigr\rangle_{ss}}{\left(\Delta-\Delta_{m}\right)+i\kappa_{m}/2},\label{a_ss}
\\\bigl\langle\hat{\sigma}_{j}\bigr\rangle_{ss}&=&\frac{\sum_{m=1}^{M}g_{m}\cos\left(\mathbf{k}_{m}\cdot\mathbf{r}_{j}\right)\bigl\langle\hat{a}_{m}\bigr\rangle_{ss}}{\Delta+i\Gamma/2}.
\label{sigma_ss}
\end{eqnarray}

Substituting the solution for $\langle \hat{\sigma}_{j} \rangle_{\text{ss}}$ into that of $\langle \hat{a}_{m} \rangle_{\text{ss}}$, we obtain coupled mode equations for $m,m^{\prime}=1, 2,\ldots, M$:
\begin{equation}
\bigl\langle\hat{a}_{m}\bigr\rangle_{ss} +\sum_{m^{\prime}\neq m}c_{mm^{\prime}}\bigl\langle\hat{a}_{m^{\prime}}\bigr\rangle_{ss}=i\Omega\delta_{m,1}, \label{aa}  \end{equation}
where we have introduced the Kronecker delta function $\delta_{m,m^{\prime}}$, and the dimensionless drive strength \begin{equation} \Omega\equiv\frac{\eta\left(i\Delta-\Gamma/2\right)}{\left[i\left(\Delta-\Delta_{1}\right)-\kappa_{1}/2\right]\left(i\Delta-\Gamma/2\right)+Ng_{1}^{2}s_{11}}. \label{omega}
\end{equation}
At this point, the definition of a (complex) mode-mode coupling matrix naturally emerges for $m^{\prime} \neq m$:
\begin{equation}
c_{mm^{\prime}} \equiv \frac{g_{m} g_{m^{\prime}} N s_{mm^{\prime}}} {\bigl[i(\Delta-\Delta_{m})-\kappa_{m}/2\bigr]\bigl(i\Delta-\Gamma/2\bigr) + N g_{m}^{2} s_{mm}},
\label{c}
\end{equation}
where the structure-factor matrix,
\begin{equation}
s_{mm^{\prime}} \equiv \frac{1}{N}\sum_{j=1}^{N}\cos\left(\mathbf{k}_{m}\cdot\mathbf{r}_{j}\right)\cos\left(\mathbf{k}_{m^{\prime}}\cdot\mathbf{r}_{j}\right),
\end{equation}
is what turns the atomic configuration into a control parameter for $c_{mm^{\prime}}$, as we shall now see.  
\begin{figure*}
\begin{centering}
\includegraphics[width=2.05\columnwidth]{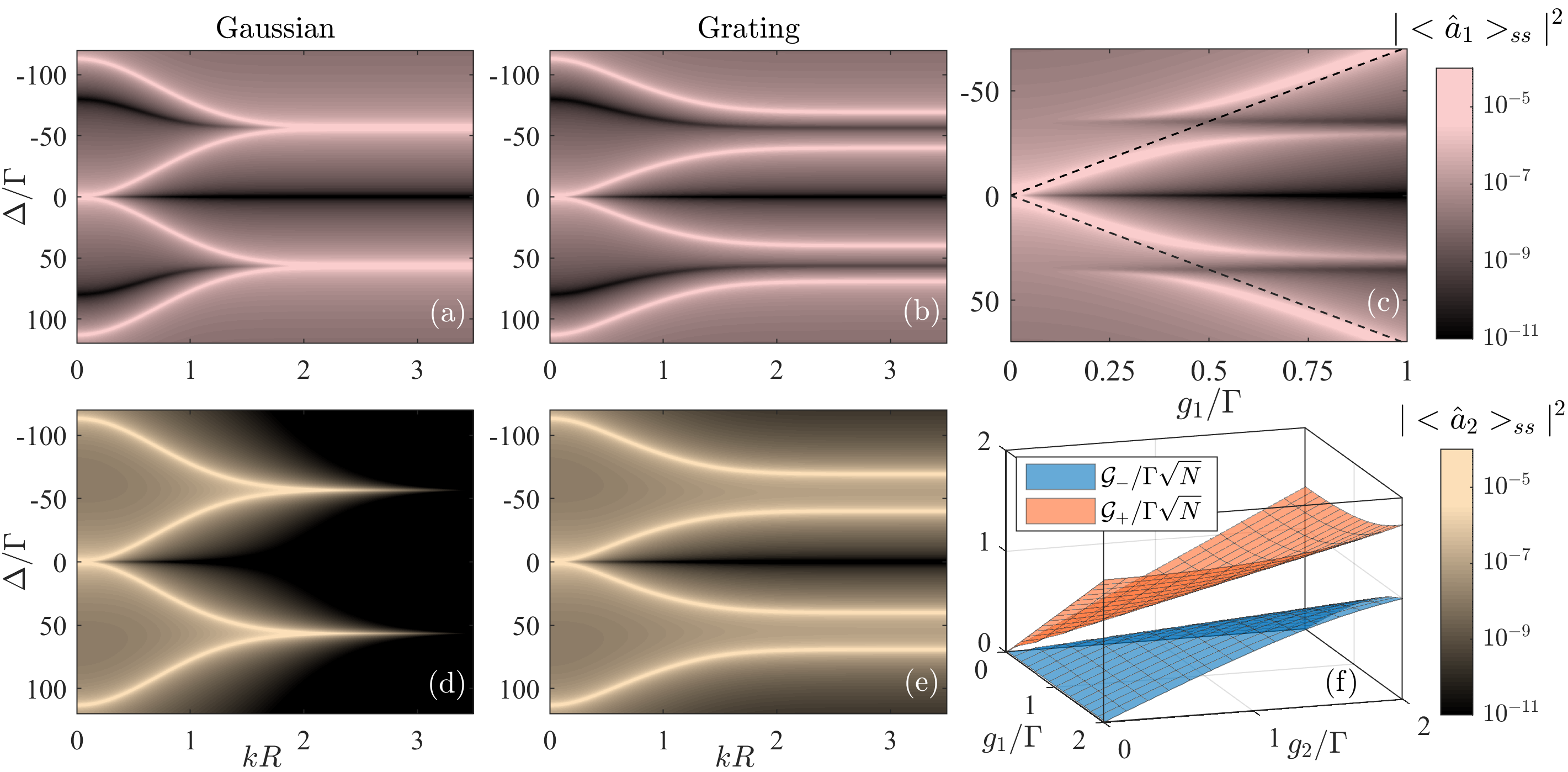}
\par\end{centering}
\caption{Amplitudes of pumped (a and b, light coral) and nonpumped (d–e, light peach) modes as functions of $\Delta/\Gamma$ and width $kR$, for pure Gaussian (first column) and grated (second column) atomic distributions. These simulations were realized for $g_1=g_2=0.8\Gamma$. (c) Amplitude of the pumped mode as a function of $\Delta$ and $g_1$, with $g_2=0.5\Gamma$. (f) Resonances $\mathcal{G}{\pm}$ as functions of $g_1$ and $g_2$. For all panels, the crossing modes are $\mathbf{k}_{1}=k(-1,1)/\sqrt{2}$ and $\mathbf{k}_{2}=k(1,1)/\sqrt{2}$, corresponding to two cavities rotated by an angle $\pi/2$ and $\mathbf{q}=(\mathbf{k}_{1}+\mathbf{k}_{2})/2=k(0,1)/\sqrt{2}$, with $\kappa_1=1\Gamma$, $\kappa_2=4\Gamma$, $\eta=0.01\Gamma$, $N=10^4$, and $\Delta_1=\Delta_2=0$.}
\label{fig2}
\end{figure*}

\section{Gaussian Cloud}

For disordered clouds with a large number of atoms, the discrete sum in the definition of $s_{mm'}$ can be replaced by the following integral over the atomic density distribution $\rho(\mathbf{r})~\cite{Bachelard2014}$:
\begin{equation}
s_{mm^{\prime}}	=	\frac{1}{N}  \int d\mathbf{r}\rho\left(\mathbf{r}\right)\cos\left(\mathbf{k}_{m}\cdot\mathbf{r}\right)\cos\left(\mathbf{k}_{m^{\prime}}\cdot\mathbf{r}\right). \label{smm}
\end{equation}
In particular, a spherically symmetric Gaussian profile of radial width $R$ yields the solution
\begin{equation}
s_{mm^{\prime}}^{\textrm{Gauss}}=\frac{1}{2}(e^{-\frac{1}{2}|\mathbf{k}_{m}-\mathbf{k}_{m^{\prime}}|^{2}R^{2}}+e^{-\frac{1}{2}\left|\mathbf{k}_{m}+\mathbf{k}_{m^{\prime}}\right|^{2}R^{2}}). \label{smm1}
\end{equation}
In the limit of a subwavelength atomic cloud, such that $R \ll 1/|\mathbf{k}_{m}+\mathbf{k}_{m'}|$, we find $s_{mm'}^{\mathrm{Gauss}} \approx 1$ for all $m,m'$, corresponding to the regime of uniform coupling between the modes. The same structure factor can be realized in a two-dimensional lattice, where all atoms are positioned near the maxima of the checkerboard standing-wave field~\cite{Donner2025}. 

As an illustration, we display in Figs.~\ref{fig2}(a) and (d) the cavity spectra for two orthogonal modes ($M=2$), $\mathbf{k}_{1}=k(-1,1)/\sqrt{2}$ and $\mathbf{k}_{2}=k(1,1)/\sqrt{2}$, and $\Delta_{2}=\Delta_{1}=0$, strongly coupled to an extremely small Gaussian cloud ($kR\rightarrow{0}$). This is a particular case of the general two-mode solution we present in Appendix A. Three resonances appear at $\Delta=\bigl\{0,\pm\sqrt{N\left(g_{1}^{2}+g_{2}^{2}\right)}\bigr\}$, with the collective nature of the splitting reflected in its dependence on the atom–mode couplings of both modes. In addition, we observe a pair of antiresonances (narrow dark lines) only in the pumped mode at $\Delta \approx \pm g_{2}\sqrt{N}$, corresponding to a pair of entangled dark states between the atoms and the nonpumped mode. 

The resonances for a subwavelength cloud can be obtained analytically from the two-mode solution of Eq.~\eqref{aa}, without resorting to the full Hamiltonian~\eqref{H} and its degeneracies, see Appendix A. They correspond to the eigenenergies of the effective reduced Hamiltonian,
\begin{equation}
\mathbf{M}_{kR\rightarrow0}^{\textrm{Gauss}}=\left(\begin{array}{ccc}
\Delta_{1} & 0 & g_{1}\sqrt{N}\\
0 & \Delta_{2} & g_{2}\sqrt{N}\\
g_{1}\sqrt{N} & g_{2}\sqrt{N} & 0
\end{array}\right), \label{Horder}
\end{equation}
which is expressed in the following single-excitation collective basis: $\ket{\psi_{c1}}=\ket{1_{1},0_{2},g_1,\ldots,g_N}$, $\ket{\psi_{c2}}=\ket{0_{1},1_{2},g_1,\ldots,g_N}$, and the symmetric superposition of atomic excitations
\begin{equation}
\ket{\psi_{kR\rightarrow 0}^{\textrm{Gauss}}}= \frac{1}{\sqrt{N}}\sum_{j=1}^{N}\ket{0_{1},0_{2},g_1,\ldots,e_{j},\ldots,g_N}.
\end{equation}
On the other hand, antiresonances and dark states originate from the drive-induced coupling between the ground state $\ket{\psi_g}=\ket{0_{1},0_{2},g_1,\ldots,g_N}$ and the single-excitation manifold. Because the weak drive does not affect the intrinsic resonances and dark states are not the focus of this work, we restrict our matrices to the single-excitation sector.
 
When the Gaussian cloud is wide enough to resolve the cavity modes, $R\gg 1/|\mathbf{k}_m-\mathbf{k}_{m'}|$ for $m\neq m'$, destructive interference suppresses the off-diagonal terms of Eq.~\eqref{smm1}, leaving $s_{mm'}^{\mathrm{Gauss}} \approx \delta_{mm'}/2$. Consequently, $c_{mm^{\prime}} \approx 0 $, and the injected light remains confined to the pumped mode, $\langle\hat{a}_{1}\rangle_{ss}\approx i\Omega$, while all other modes are not populated ($\langle\hat{a}_{m}\rangle_{ss}\approx0$ for $m\neq1$). Thus, a sufficiently extended Gaussian cloud reproduces the case of a homogeneous distribution interacting with a single cavity mode~\cite{Birgitta2004,Domokos2025}, and is therefore inefficient at coupling different modes. This behavior becomes evident in the panels of Figs.~\ref{fig2}(a) and (d) as $kR$ increases: the pumped-mode spectrum rapidly converges to a pair of single-mode Lorentzian peaks, at $\Delta=\pm g_{1}\sqrt{N/2}$, while the amplitude of the nonpumped mode decays exponentially. Moreover, in mode 1, a single antiresonance remains at $\Delta=0$, revealing the formation of a purely atomic dark state induced by the weak pump~\cite{Emary2013}.

The resonances of the effective single-mode system, which are generated by a large Gaussian cloud, can be fully obtained from the following $2\times2$ matrix:
\begin{equation}
\mathbf{M}_{kR\rightarrow\infty}^{\textrm{Gauss}}=\left(\begin{array}{cc}
\Delta_{1} & g_{1}\sqrt{\frac{N}{2}}\\
g_{1}\sqrt{\frac{N}{2}} & 0
\end{array}\right). \label{Hgauss} 
\end{equation} 
Eq.\eqref{Hgauss} represents the Tavis-Cummings Hamiltonian in a basis composed of the single-photon state related to the pumped mode $\ket{\psi_{c1}}$, and a collective atomic excitation following the cavity profile:
\begin{equation}
\ket{\psi_{kR\rightarrow\infty}^{\textrm{Gauss}}}=\sum_{j=1}^{N}\frac{\cos\left(\mathbf{k}_{1}\cdot\mathbf{r}_{j}\right)}{\sqrt{N/2}}\bigl|0_{1},0_{2},g_{1},\ldots,e_{j},\ldots g_{N}\bigr\rangle,
\end{equation}
see Appendix B. Thus, we are reducing the complexity of the many-body system by one dimension compared to that of a subwavelength atomic cloud, and modifying drastically the hybridization of the atom-photon system.

\section{Grated cloud} Still in the context of Gaussian distributions, Figs.~\ref{fig2}(a) and (d) reveal an intermediate regime ($0 < kR \lesssim 2$) where the spectra acquire a more intricate structure. As the cloud shrinks, the gradual recovery of mode–mode coupling causes the two original symmetric peaks to further split, giving rise to four bright states and three accompanying dark states. However, because the structure factor decays exponentially with $kR$ [see Eq.~\eqref{smm1}], the cloud widths that sustain this rich spectrum, keeping the collective strong-coupling regime, are much narrower than those typically achievable in cold-atom platforms. This highlights the importance of exploring alternative atomic configurations that can support the full spectral structure under more accessible conditions.

Let us consider a Gaussian density distribution with a spatial modulation of wavevector $\mathbf{q}$,
\begin{equation}
\rho\left(\mathbf{r}\right)=\frac{N\cos^{2}\left(\mathbf{q}\cdot\mathbf{r}\right)e^{-\frac{1}{2}\frac{\left|\mathbf{r}\right|^{2}}{R^{2}}}}{\sqrt{2\pi^{3}}R^{3}\left(1+e^{-2\left|\mathbf{q}\right|^{2}R^{2}}\right)}, \label{rhoc}
\end{equation}
which describes a grating~\cite{Donner2017,Zhao2018} with a Gaussian envelope. Such an atomic distribution can be readily implemented experimentally by applying a one-dimensional optical lattice, with the associated structure factor integral~\eqref{smm} admitting the following exact solution:
\begin{multline}
 s_{mm^{\prime}}^{\textrm{Grated}}=\frac{1}{4\left(1+e^{-2\left|\mathbf{q}\right|^{2}R^{2}}\right)}\left(4s_{mm^{\prime}}^{\textrm{Gauss}}+\right.\\e^{-\frac{1}{2}\left|2\mathbf{q}+\mathbf{k}_{m}+\mathbf{k}_{m^{\prime}}\right|^{2}R^{2}}+e^{-\frac{1}{2}\left|2\mathbf{q}-\mathbf{k}_{m}+\mathbf{k}_{m^{\prime}}\right|^{2}R^{2}}+\\\left.e^{-\frac{1}{2}\left|2\mathbf{q}+\mathbf{k}_{m}-\mathbf{k}_{m^{\prime}}\right|^{2}R^{2}}+e^{-\frac{1}{2}\left|2\mathbf{q}-\mathbf{k}_{m}-\mathbf{k}_{m^{\prime}}\right|^{2}R^{2}}\right). \label{smm2} 
\end{multline}  
Note that the pure Gaussian result is recovered for $\mathbf{q}=0$. Applying the limit $kR\rightarrow{0}$ in Eq.~\eqref{smm2}, one recovers $s_{mm^{\prime}}^{\textrm{Grated}}=s_{mm^{\prime}}^{\textrm{Gauss}} \approx 1$ for all $\mathbf{q}$, as the grating becomes irrelevant. However, for extended clouds, a vector $\mathbf{q}$ that meets any of the constructive interference conditions, $\left| 2\mathbf{q} \pm \left(\mathbf{k}_m+\mathbf{k}_{m^{\prime}}\right) \right| \ll 1/R$ or $\left| 2\mathbf{q} \pm \left(\mathbf{k}_m-\mathbf{k}_{m^{\prime}}\right) \right| \ll 1/R$, prevents the elements of the structure factor matrix from vanishing: $s_{mm^{\prime}}^{\textrm{Grated}} \approx 1/4$ for $m^{\prime} \neq m$ and $s_{mm} \approx  1/2$. This occurs because a suitable Fourier component of the grating provides a Bragg condition that allows light scattering between a given pair $\left\{ m,m^{\prime}\right\} $ of cavity modes~\cite{Courteille2006}.

We illustrate this point in Figs.~\ref{fig2}(b) and~\ref{fig2}(e), which display the cavity spectra for the same two modes as in Figs.~\ref{fig2}(a) and~\ref{fig2}(d), but now in the presence of an atomic grating of wave vector $\mathbf{q}=(\mathbf{k}_{1}+\mathbf{k}_{2})/2$. In contrast to the pure Gaussian case, the amplitude of the nonpumped mode no longer vanishes with increasing $kR$, and the resonances in both modes do not merge into two peaks. Instead, they remain distinct, with their frequencies drawn closer together due to the reduced structure factor ($s_{12}^{\textrm{Grated}} \approx 1/4$). For the two modes of Fig.~\ref{fig2}, a linear deviation $\mathbf{q} = (1 + \epsilon)\left(\mathbf{k}_1+\mathbf{k}_{2}\right)/2$ from the optimal coupling condition $\epsilon=0$, with $\epsilon$ a nonnegative parameter,  results in the asymptotic expansion $s_{12^{\prime}}^{\textrm{Grated}}\approx e^{-\left(\epsilon kR\right)^{2}}/4$ for large $kR$. Therefore, the constructive interference condition remains robust for $\epsilon\ll1/kR$.

For a sufficiently extended grating satisfying one of the four conditions mentioned before, and still under the strong-coupling regime, the four resonances correspond to $\Delta=\bigl\{\mathcal{G}_{\pm},-\mathcal{G}_{\pm}\bigr\}$, where we introduce the following collective frequencies:
\begin{equation}
\mathcal{G}_{\pm}\equiv\frac{1}{2}\sqrt{N\left(g_{1}^{2}+g_{2}^{2}\pm\sqrt{\left(g_{1}^{2}+g_{2}^{2}\right)^{2}-3g_{1}^{2}g_{2}^{2}}\right)}.
\end{equation}
Note that the cross term $3g_{1}g_{2}$ is a signature of the collective mode–mode coupling. According to Fig.~\ref{fig2}(f), the subsplittings satisfy $\mathcal{G}{+}\geq \mathcal{G}{-}\geq 0$, with their separation scaling as $\mathcal{G}_{+}-\mathcal{G}_{-}=\sqrt{N\left(g_{1}^{2}+g_{2}^{2}-\sqrt{3}g_{1}g_{2}\right)/2}$ for $g_{1},g_{2}>0$. In addition, Fig.~\ref{fig2}(c) shows how a nonzero $g_{2}$ modifies the single-cavity spectrum (dashed lines), generating two additional avoided crossings separated by antiresonances. Thus, the atomic grating preserves the collectivity that manifests itself as a function of distinct atom–mode couplings, a feature that is unattainable with homogeneous clouds. 

As previously mentioned, a suitable grated distribution preserves three distinct antiresonances in large cloud limit, $kR \to \infty$. The central antiresonance, at $\Delta=0$, corresponds to a dark state formed by an antisymmetric superposition of the ground state and all elements of the single-excitation basis with no photons in either mode~\cite{Emary2013}. In this case, the excitation tends to remain trapped within the atomic ensemble, which explains the antiresonance in both modes. By contrast, the side antiresonances at $\Delta=\pm g_{2}\sqrt{N/2}$ ($\pm g_{2}\sqrt{N}$ in the subwavelength case) correspond to hybridized atom–photon dark states that involve a photon in the second mode. Since the weak drive cannot simultaneously populate both cavities, destructive interference ensures that the photon is confined to the nonpumped mode only. This mechanism can be exploited to achieve selective photon population through density modulation in the atomic cloud.

The resonances of an effective two-mode system sustained by the grating atomic distribution are the eigenvalues of the following $4 \times 4$ matrix:
\begin{equation}
\mathbf{M}_{kR\rightarrow\infty}^{\textrm{Grated}}=\left(\begin{array}{cccc}
\Delta_{1} & 0 & \frac{g_{1}}{2}\sqrt{\frac{N}{2}} & \frac{g_{1}}{2}\sqrt{\frac{3N}{2}}\\
0 & \Delta_{2} & -\frac{g_{2}}{2}\sqrt{\frac{N}{2}} & \frac{g_{2}}{2}\sqrt{\frac{3N}{2}}\\
\frac{g_{1}}{2}\sqrt{\frac{N}{2}} & -\frac{g_{2}}{2}\sqrt{\frac{N}{2}} & 0 & 0\\
\frac{g_{1}}{2}\sqrt{\frac{3N}{2}} & \frac{g_{2}}{2}\sqrt{\frac{3N}{2}} & 0 & 0
\end{array}\right). \label{Hgrat}
\end{equation}
The associated basis now includes the states of the subwavelength case with all atoms in the ground state, together with the two collective atomic states
\begin{flalign}
\ket{-}&=\sum_{j=1}^{N}\frac{\cos\left(\mathbf{k}_{1}\cdot\mathbf{r}_{j}\right)-\cos\left(\mathbf{k}_{2}\cdot\mathbf{r}_{j}\right)}{\sqrt{N/2}}\bigl|0_{1},0_{2},g,\ldots,e_{j},\ldots g\bigr\rangle,\\\ket{+}&=\sum_{j=1}^{N}\frac{\cos\left(\mathbf{k}_{1}\cdot\mathbf{r}_{j}\right)+\cos\left(\mathbf{k}_{2}\cdot\mathbf{r}_{j}\right)}{\sqrt{3N/2}}\bigl|0_{1},0_{2},g,\ldots,e_{j},\ldots g\bigr\rangle.
\end{flalign}
Hence, the effective system dimension for two modes coupled to an extended grating is larger than that of the subwavelength case, as the degeneracy of the full Hamiltonian is lifted when atoms sample the two-mode field profile.

When considering longer timescales, the atomic center-of-mass motion can no longer be neglected. In this regime, inhomogeneous Doppler broadening and momentum diffusion introduce additional dephasing channels, thereby enhancing the overall decoherence of the system. The velocity-dependent frequency shifts lead to a progressive loss of phase synchronization among the atomic dipoles, which impacts interference-based collective phenomena. In particular, the effective mode–mode coupling mediated by the atomic ensemble, discussed here, is gradually suppressed as motional dephasing increases. Consequently, the corresponding frequency subsplittings in the system spectrum become progressively less pronounced, eventually approaching the unresolved limit for sufficiently strong Doppler broadening.  In addition, depending on detuning, dipole forces will act on the atoms, leading to the formation of self-consistent potentials as known from many-body cavity QED experiments.

\section{Conclusion} 

We have shown that spatial modulations in atomic ensembles offer a flexible means of tailoring the coupling between modes of different cavities. Large Gaussian clouds suppress these couplings through destructive interference, whereas deep subwavelength clouds restore the uniform-coupling limit. To preserve the multimode atom-photon hybridization for extended clouds, such as dark states relevant for a range of applications~\cite{Vanier2005,Fleischhauer2016,Esslinger2022}, we proposed imprinting a grating onto the Gaussian envelope of the distribution. This kind of spatial inhomogeneity enables selective photon exchange between a given pair of cavities, in close analogy with Bragg scattering in crystals, and generates subsplittings that directly reveal the strength of mode–mode coupling.

Further spatial modulations of the atomic distribution can extend this mechanism to multiple cavity modes, with additional Fourier components of the density profile that preserve the coupling between several pairs of modes. As a consequence, the spectrum may exhibit multiple frequency subsplittings, enabling the selective population of different optical cavity modes via the formation of multimode bright and dark collective states. Also, the inclusion of two or more atomic excitations induces nonlinear mode–mode interactions, which may produce counterintuitive spectral subsplittings that could serve as fingerprints of many-body quantum correlations between modes. Finally, our approach opens promising avenues for finely engineering the dimensionality of effective many-body systems, where large grating clouds are instrumental in achieving this goal.

\section{Acknowledgment}

The authors thank Alexander Baumgärtner, Davide Dreon, and Simon Hertlein for stimulating discussions. C.E.M. and R.B. acknowledge funding from the S\~ao Paulo Research Foundation (FAPESP, Grants Nos. 2022/00209-6, 2023/11118-4, 2023/01213-0 and 2023/03300-7) and from the Brazilian National Council for Scientific and Technological Development (CNPq, Grants Nos. 402660/2019-6, 201765/2020-9, 151895/2022-8, 158497/2023-6, 443989/2024-9 and 315107/2025-1), and by the Swiss National Science Foundation (SNSF), Projects No. IZBRZ2-186312/1  223274 and 217124. This project is funded within the QuantERA II Programme, which has received funding from the EU's Horizon 2020 research and innovation programe under Grant Agreement No. 101017733, as well as from the funding organizations SNSF under Grant No. 221538. T.D. further acknowledges funding from the Swiss State Secretariat for Education, Research and Innovation (SERI) under Grant No. MB22.00090.

\appendix

\section{Resonances for the case of a subwavelength atomic distribution}

For the case of two modes only, the general steady-state solution for Eq. \eqref{aa} reads:
\begin{widetext}
\begin{eqnarray}
\bigl\langle\hat{a}_{1}\bigr\rangle_{ss}&=&\frac{-i\eta\left(i\Delta-\frac{\Gamma}{2}\right)\left[g_{2}^{2}Ns_{22}+\left(i\Delta-\frac{\Gamma}{2}\right)\left(i\left(\Delta-\Delta_{2}\right)-\frac{\kappa_{2}}{2}\right)\right]}{g_{1}^{2}g_{2}^{2}N^{2}s_{12}s_{21}-\left[g_{1}^{2}Ns_{11}+\left(i\Delta-\frac{\Gamma}{2}\right)\left(i\left(\Delta-\Delta_{1}\right)-\frac{\kappa_{1}}{2}\right)\right]\left[g_{2}^{2}Ns_{22}+\left(i\Delta-\frac{\Gamma}{2}\right)\left(i\left(\Delta-\Delta_{2}\right)-\frac{\kappa_{2}}{2}\right)\right]},\label{a1_general}
\\\bigl\langle\hat{a}_{2}\bigr\rangle_{ss}&=&\frac{i\eta g_{1}g_{2}Ns_{21}\left(i\Delta-\frac{\Gamma}{2}\right)}{g_{1}^{2}g_{2}^{2}N^{2}s_{12}s_{21}-\left[g_{1}^{2}Ns_{11}+\left(i\Delta-\frac{\Gamma}{2}\right)\left(i\left(\Delta-\Delta_{1}\right)-\frac{\kappa_{1}}{2}\right)\right]\left[g_{2}^{2}Ns_{22}+\left(i\Delta-\frac{\Gamma}{2}\right)\left(i\left(\Delta-\Delta_{2}\right)-\frac{\kappa_{2}}{2}\right)\right]}, \label{a2_general}
\end{eqnarray}
\end{widetext}
which is the solution used to generate all density plots. In the collective strong-coupling regime ($\Gamma \approx \kappa_1 \approx \kappa_2 \approx 0$), the numerators and denominators of Eqs.~\eqref{a1_general} and~\eqref{a2_general} become real polynomials whose orders depend on the structure-factor matrix. Their roots correspond, respectively, to the antiresonance and resonance conditions of the lossless system.

We now specify in Eqs.~\eqref{a1_general} and~\eqref{a2_general} the structure factor matrix for a subwavelength atomic cloud ($s_{mm^{\prime}} \approx 1$ for $m,m^{\prime}=1,2$). In this regime, the resonances discussed in the main text are obtained as the roots of the following third-order polynomial:
\begin{multline}
\Delta^{3}-\Delta^{2}\left(\Delta_{1}+\Delta_{2}\right)-\Delta\left[\left(g_{1}^{2}+g_{2}^{2}\right)N-\Delta_{1}\Delta_{2}\right]\\+\left(g_{1}^{2}\Delta_{2}+g_{2}^{2}\Delta_{1}\right)N=0. \label{res_sub}
\end{multline}
The antiresonances, by contrast, arise exclusively from the numerator of the solution for mode 1 (Eq.~\eqref{a2_general}) and correspond to the roots of the second-order polynomial $\Delta\left(\Delta-\Delta_{2}\right)-g_{2}^{2}N=0$. In the main text, we present the analytical solutions for these polynomials for the degenerate case ($\Delta_2=\Delta_1=0$).

To determine the corresponding single-excitation state basis, we must note that the spatial effect is negligible for a subwavelength cloud, and that both modes excite all atoms with equal weights ($\cos\left(\mathbf{k}_{1}\cdot\mathbf{r}_{j}\right) = \cos\left(\mathbf{k}_{2}\cdot\mathbf{r}_{j}\right) \approx1$). Thus, the atomic component of this basis, $\ket{\psi_{kR\rightarrow 0}^{\textrm{Gauss}}}$, can only be a uniform superposition of single-excitation states, independent of the atomic positions, as presented in the main text. Under these assumptions, we can obtain all matrix elements of $\hat{H}$, neglecting two-excitation contributions: $\mathbb{1} \Delta \left(N/2-1\right) +\mathbf{M}_{kR\rightarrow0}^{\textrm{Gauss}}$, where $\mathbb{1}$ denotes the identity matrix. Since the global trivial shift $\Delta \left(N/2-1\right)$ does not affect the dynamical solutions, we omit it to obtain the characteristic equation of $\mathbf{M}_{kR\rightarrow0}^{\textrm{Gauss}}$ that exactly matches Eq.~\eqref{res_sub}.

\section{Resonances for the case of a large Gaussian atomic distribution}

Using now the structure factor matrix for a large Gaussian cloud, where $s_{11}=s_{22}\approx1/2$ and $s_{12}=s_{21}\approx0$, and applying again the strong-coupling regime ($\Gamma\approx\kappa_1\approx0$), we obtain the following solution for the modes amplitudes:
\begin{equation}
\bigl\langle\hat{a}_{1}\bigr\rangle_{ss}=\frac{\eta\Delta}{\Delta\left(\Delta-\Delta_{1}\right)-g_{1}^{2}N/2},\quad\bigl\langle\hat{a}_{2}\bigr\rangle_{ss}\approx0 \label{a_gauss}
\end{equation}
The roots of the denominator $\Delta\left(\Delta-\Delta_{1}\right)-g_{1}^{2}N/2=0$ determine the relevant resonance conditions which were presented in the main text. Meanwhile, the numerator indicates the absence of photons at $\Delta=0$ (antiresonance). Since we are in the weak-drive regime, and the nonpumped mode receives no photons, the corresponding effective Hamiltonian should be determined solely by the following subspace:
\begin{equation}
\Bigl\{\bigl|1_{1},0_{2},g_{1},\ldots,g_{N}\bigr\rangle,\sum_{j=1}^{N}\alpha_{j}\bigl|0_{1},0_{2},g,\ldots,e_{j},\ldots g\bigr\rangle\Bigr\}, \label{basis}
\end{equation}
where the coefficients $\alpha_j$ remain unknown.

Next, we compute the matrix elements of $\hat{H}$ in the basis~\eqref{basis}. In particular, we emphasize the action of the full Hamiltonian on the superposition component:
\begin{multline}
\hat{H}\sum_{j}\alpha_{j}\bigl|0_{1},0_{2},g_{1},\ldots,e_{j},\ldots g_{N}\bigr\rangle \\=\eta\sum_{j}\alpha_{j}\bigl|1_{1},0_{2},g_{1},\ldots,e_{j},\ldots g_{N}\bigr\rangle\\+\Delta\left(\frac{N}{2}-1\right)\sum_{j}\alpha_{j}\bigl|0_{1},0_{2},g_{1},\ldots,e_{j},\ldots g_{N}\bigr\rangle\\+g_{1}\left(\sum_{j=1}^{N}\alpha_{j}\cos\left(\mathbf{k}_{1}\cdot\mathbf{r}_{j}\right)\right)\bigl|1_{1},0_{2},g_{1},\ldots,g_{N}\bigr\rangle\\+g_{2}\left(\sum_{j=1}^{N}\alpha_{j}\cos\left(\mathbf{k}_{2}\cdot\mathbf{r}_{j}\right)\right)\bigl|0_{1},1_{2},g_{1},\ldots,g_{N}\bigr\rangle. \label{Hbeta}
\end{multline}
where all two-excitation states have been neglected, as they do not contribute in the weak-drive regime. Considering that Gaussian clouds do not allow light scattering into the nonpumped mode, and $s_{11}\approx1/2$ and $s_{12}\approx0$, we can infer $\alpha_j=\cos\left(\mathbf{k}_{1}\cdot\mathbf{r}_{j}\right)/\sqrt{N/2}$. 
Applying the result above in Eq.~\eqref{Hbeta}, we find the effective matrix representation $\mathbf{M}_{kR\rightarrow \infty}^{\textrm{Gauss}}$, up to the same energy shift $\Delta \left(N/2-1\right)$. The corresponding eigenvalue equation of this matrix coincides exactly with the denominator of the mode-1 amplitude in~\eqref{a_gauss}, as we were looking for. 

\section{Resonances for the case of a grated atomic distribution}

A similar procedure can be followed to obtain the Hamiltonian matrix $\mathbf{M}_{kR\rightarrow \infty}^{\textrm{Grated}}$ for the grated cloud. In this case, the key difference is that the spatial profiles of the two modes must now be retained in the definition of the state basis. The effective matrix can therefore be derived from the action of the Hamiltonian~\eqref{H} on the following arbitrary single-excitation state:
\begin{multline}
\ket{\pm}=\sum_{j=1}^{N}\left[\alpha_{j}^{\pm}\cos\left(\mathbf{k}_{1}\cdot\mathbf{r}_{j}\right)+\beta_{j}^{\pm}\cos\left(\mathbf{k}_{2}\cdot\mathbf{r}_{j}\right)\right]\\\times\bigl|0_{1},0_{2},g,\ldots,e_{j},\ldots g\bigr\rangle.
\end{multline}

\bibliography{ref}
\end{document}